\documentclass[]{jjap3}

\usepackage{amsmath}
\usepackage{amssymb}
\usepackage{color}
\usepackage{graphicx}
\usepackage{bm}
\usepackage{txfonts}
\usepackage{tikz}
\usepackage{pgffor}
\usepackage{verbatim}
\usepackage{pstricks}
\usepackage{epstopdf}

\DeclareFontFamily{U}{euc}{}
\DeclareFontShape{U}{euc}{m}{n}{<-6>eurm5<6-8>eurm7<8->eurm10}{}
\DeclareSymbolFont{AMSc}{U}{euc}{m}{n}
\DeclareMathSymbol{\umu}{\mathord}{AMSc}{"16}

\title{Temperature profile of the Thomson-effect-induced heat release/absorption in junctionless single conductors}

\author{Takahiro Chiba$^1$, Ryo Iguchi$^2$, Takashi Komine$^3$, Yasuhiro Hasegawa$^4$, and Ken-ichi Uchida$^{2,5}$}
\inst{$^1$National Institute of Technology, Fukushima College, Iwaki, Fukushima 970-8034, Japan \\
$^2$National Institute for Materials Science, Tsukuba, Ibaraki 305-0047, Japan \\
$^3$Graduate School of Science and Engineering, Ibaraki University, Hitachi, Ibaraki 316-8511, Japan \\
$^4$Graduate School of Science and Engineering, Saitama University, Saitama, Saitama 338-8570, Japan \\
$^5$Institute for Materials Research, Tohoku University, Sendai, Miyagi 980-8577, Japan} 

\abst{The Thomson effect induces heat release or absorption under the simultaneous application of a charge current and a temperature gradient to conductors. Here, we theoretically investigate the temperature profile due to the Thomson-effect-induced heat release/absorption in junctionless single conductors which can be a simple temperature modulator. We also perform analysis of the temperature profile for realistic conductors. As a result, we find that, for a conductor with a large Thomson coefficient, the temperature derivative of the Seebeck coefficient, the Thomson-effect-induced heat absorption overcomes the Joule heating, resulting in current-induced cooling in the bulk region. We also elucidate that a feedback effect of the Thomson effect stabilizes the system temperature to one-side of the heat bath, which reflects the fact that the Thomson effect is dependent on the position and proportional to the local temperature gradient. This work will be the basis for thermal management utilizing the Thomson effect.
}

\begin{document}
\maketitle

\section{Introduction}

The linear-response thermoelectric phenomena, i.e., the Seebeck and Peltier effects, have been studied in the fundamental physics for more than a century and also used in a wide range of applications including energy harvesting and thermal management \cite{Rowe95,Bell08,Goupil11,Liu15,Li21}. A Seebeck power generator that is interpreted as a thermoelectric engine converts heat into electrical work. Also, a Peltier temperature modulator (Peltier device) corresponds to heat pump of the electron  gas (or the inverse cycle of the thermoelectric engine). 
Unlike the Peltier device with the complicated cascade structure, it is expected that a thermoelectric device operated by the Thomson effect works as a simple temperature modulator without the construction of junction structures.
The Thomson effect, one of the nonlinear thermoelectric phenomena, induces heat release or absorption under the simultaneous application of a charge current density ${\bf J}$ and a temperature ($T$) gradient $\bm{\nabla}T$ to conductors. The induced heat production rate per unit volume is given by
\begin{align}
\dot{q} = -\tau(T){\bf J}\cdot\bm{\nabla}T
,\label{Thomson q}
\end{align}
where 
\begin{align}
\tau(T) = T\frac{dS}{dT}
\label{Thomson coefficient}
\end{align}
is the Thomson coefficient and $S$ the Seebeck coefficient. Equation~(\ref{Thomson coefficient}) indicates that conductors with a sharp $T$ dependence of $S$ are potential candidates to generate the large Thomson effect. However, for many conventional thermoelectric materials, the Thomson-effect-induced heat release/absorption is small so that it has not been focused in the past studies.

Recently, Modak \textit{et.~al.} found that due to the steep $T$ dependence of the Seebeck coefficient, FeRh-based alloys show huge values of the Thomson coefficient approaching $\sim 1000$~$\umu$V/K around room temperature \cite{Modak21}. They demonstrated that the Thomson-effect-induced cooling can be larger than the Joule heating in the FeRh-based alloy in a steady state. Furthermore, Uchida \textit{et.~al.} reported that the magnitude of the Thomson-effect-induced temperature change in Bi$_{88}$Sb$_{12}$ is strongly enhanced by applying a magnetic field \cite{Uchida20}. Thus, control of the local temperature by the Thomson effect might be a potential technology for thermoelectrics so that it is important to properly investigate the temperature distribution induced by the Thomson effect.
On the theoretical side, a thermoelectric cooler that incorporates the contribution of the Thomson term into the Peltier device has been investigated by using the compatibility approach \cite{Snyder12,Seifert13}. Although the influence of the Thomson effect in a Peltier device has been theoretically discussed \cite{Snyder12,Seifert13,Thiebaut17,Feng18,Bubanja22}, a temperature modulation driven solely by the Thomson effect in a single conductor without heterojunctions has not been investigated so far. Since as seen in Eq.~(\ref{Thomson q}) the Thomson effect is driven by the temperature gradient itself, the Thomson-effect-induced heat release/absorption gives feedback to its own output. Hence, the detailed analysis of its temperature profile is essential for the thermal management based on the Thomson effect in junctionless single conductors.

In this paper, we theoretically study the temperature profile of the Thomson-effect-induced heat release/absorption in junctionless single conductors which can be a simple temperature modulator. For a conductor with a large Thomson coefficient, we show that the Thomson-effect-induced heat absorption overcomes the Joule heating, leading to current-induced cooling in the bulk region from the initial state without the charge current. We also elucidate how a feedback of the Thomson effect influences a local temperature in the system. We give a more realistic discussion on the temperature modulation of the junctionless single conductors with material parameters.


\section{Model}\label{model}

\begin{figure}[ptb]
\begin{centering}
\includegraphics[width=0.45\textwidth,angle=0]{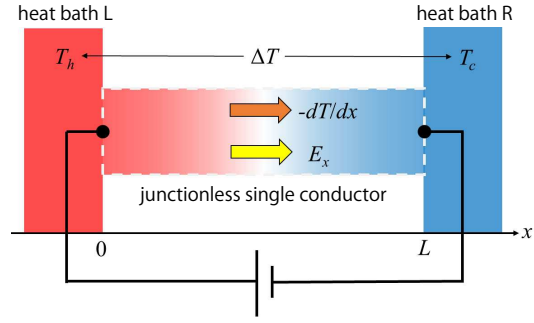} 
\par\end{centering}
\caption{
Schematic of a junctionless single conductor in the one-dimensional geometry in which $T_{\rm h}~(>T_{\rm c})$ and $T_{\rm c}$ denote the hot- and cold-side temperatures by assuming heat bath (L and R) contacts, respectively. Current-induced cooling in the bulk region from the initial state with the uniform temperature gradient occurs when a heat absorption by the Thomson effect is larger than the Joule heating. The detail of the operation is described in the text.
}
\label{fig:system}
\end{figure}

We consider a junctionless single conductor (with the cross-sectional area $A$ and the length $L$) connected with a perfect conductor wiring at both the ends and having an external  temperature difference $\Delta T = T_{\rm h} - T_{\rm c}$ with $T_{\rm h}$ and $T_{\rm c}$ being the hot- and cold-side temperatures, as shown in Fig.~\ref{fig:system}. Without loss of generality, we restrict thermoelectric transports to one spatial dimension. In the presence of an electric field ($E_x$) and a temperature gradient ($-dT/dx$) along the $x$-axis, charge and heat current densities are respectively described as,   
\begin{align}
J &= \sigma(T)E_x - \sigma(T)S(T)\frac{dT}{dx}
,\label{J}\\
J_Q &= S(T)TJ - \kappa(T)\frac{dT}{dx}
,\label{JQ}
\end{align}
where $\sigma(T) = \rho^{-1}(T)$ is the electrical conductivity as an inverse of the electrical resistivity and $\kappa(T)$ the thermal conductivity. 
In the steady state, i.e., $-d\left(  J_Q + \mu_eJ/e\right)/dx = 0$ with $e$ being the elemental charge and $\mu_e$ the electrochemical potential, the local energy balance over the junctionless single conductor is given by \cite{CallenBook}
\begin{align}
\frac{d}{dx}\left(  - \kappa(T)\frac{dT}{dx}\right)
= \rho(T)J^2 - \tau(T)\frac{dT}{dx}J
,\label{balance00}
\end{align}
where we assume that the Seebeck coefficient is only dependent on location $x$ via temperature, i.e. an uniform material is assumed. In contrast to the Joule heating that is proportional to $J^2$, the Thomson term is proportional to $J$, enabling the control of heat release or absorption depending on whether a charge current flows in the direction parallel or antiparallel to a temperature gradient. Importantly, the sign and magnitude of the Thomson term are determined by $\tau(T)$. 

\section{Discussion}\label{Discussion}

In this section, we show how a local temperature $T(x)$ of a junctionless single conductor depend on the thermoelectric coefficients. 
We introduce a system temperature difference $\Delta T_{\rm system}$ to characterize the Thomson-effect-induced bulk temperature modulation. 
We also perform analysis of the temperature profile for realistic conductors.

\subsection{Local temperature $T(x)$ of the junctionless single conductor}

Here, we investigate how the local temperature $T(x)$ of the junctionless single conductor depends on the transport coefficients by solving Eq.~(\ref{balance00}). Assuming a weak $T$ dependence of the thermal conductivity ($d\kappa/dT \approx 0$) and $\Delta T \ll T$, we replace Eq.~(\ref{balance00}) by the following heat equation with temperature averaged thermoelectric coefficients denoted by $\langle C \rangle_T = \left(  1/\Delta T\right)\int_{T_{\rm c}}^{T_{\rm h}} C(T) dT$~$(C = \rho, \kappa, \tau)$:
\begin{align} 
-\langle \kappa \rangle_T\frac{d^2T}{dx^2} + \langle \rho \rangle_TJ^2 - \langle \tau \rangle_T\frac{dT}{dx}J = 0
.\label{balance pthw}
\end{align}
This heat equation describes the local energy balance among the Fourier (parameterized by $\langle \kappa \rangle_T$), Joule, and Thomson heats.
Hereafter, we introduce a charge current $I = AJ$, the electrical resistance $R = \langle \rho \rangle_TL/A$, and the thermal conductance $K = \langle \kappa \rangle_TA/L$ on Eq.~(\ref{balance pthw}), which gives
\begin{align}
T(x) = T_{\rm h}  + \frac{RI}{\langle \tau \rangle_T}\frac{x}{L}
- \left(  \Delta T + \frac{RI}{\langle \tau \rangle_T}\right)\frac{ \exp{\left(  \frac{\langle \tau \rangle_TI}{K}\frac{x}{L}\right)}- 1}{\exp{\left(  \frac{\langle \tau \rangle_TI}{K}\right)} - 1}
.\label{T(x)}
\end{align}
Here, we set the thermal boundary condition as $T(x=0) = T_{\rm h} = \Delta T + T_{\rm c}$ and $T(x=L) = T_{\rm c}$ in Fig.~\ref{fig:system}. 

\begin{figure}[ptb]
\begin{centering}
\includegraphics[width=0.45\textwidth,angle=0]{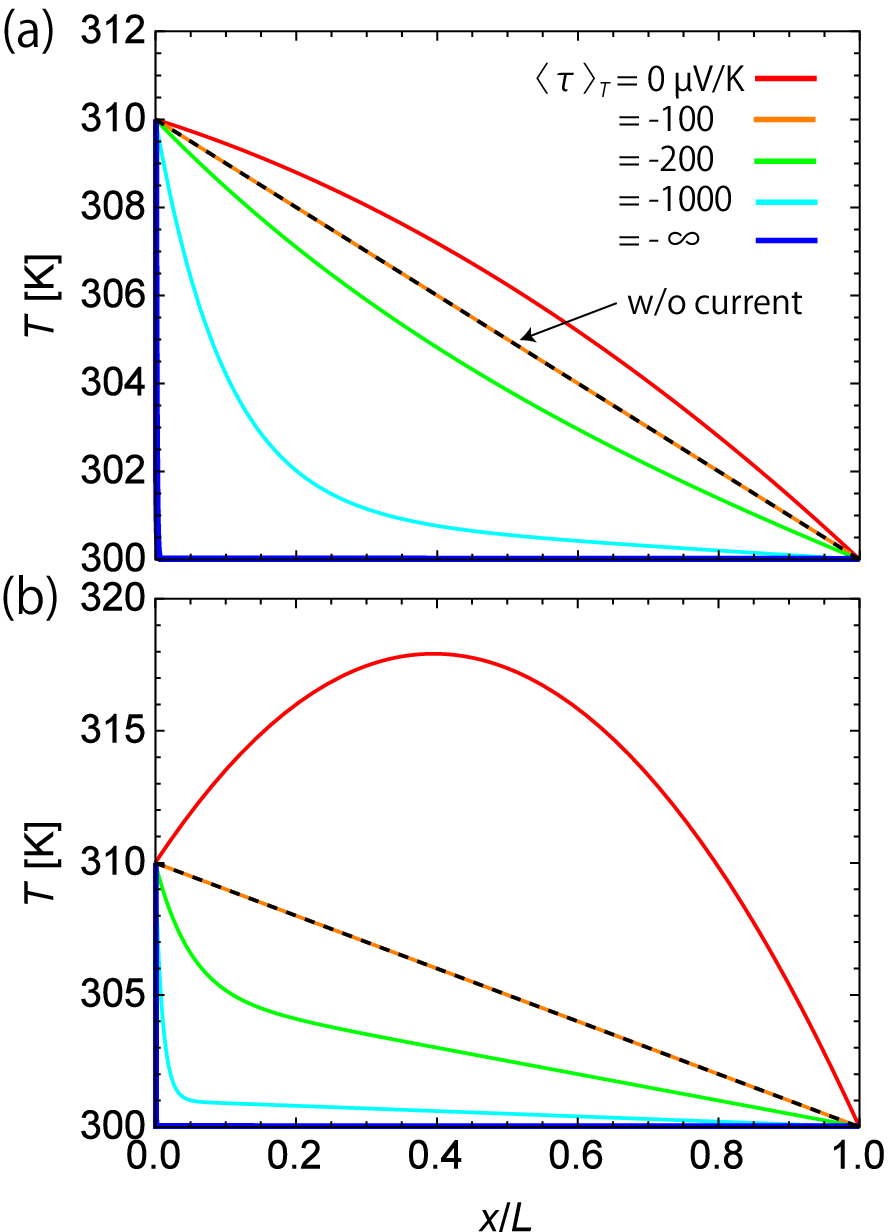} 
\par\end{centering}
\caption{Local temperature due to the Thomson-effect-induced heat absorption as a function of a normalized position $x/L$ for different values of $\langle \tau \rangle_T$~[$\umu$V/K] with (a) $K = 10$~$\umu$W/K and (b) $K = 1$~$\umu$W/K. The situation in the absence of current is described by dashed line. In these plots, $R = 0.01~\Omega$, $I = 0.1$~A, $T_{\rm c} = 300$~K, and $\Delta T = 10$~K are used. 
 }
\label{fig:T(x)c}
\end{figure}

\begin{figure}[ptb]
\begin{centering}
\includegraphics[width=0.45\textwidth,angle=0]{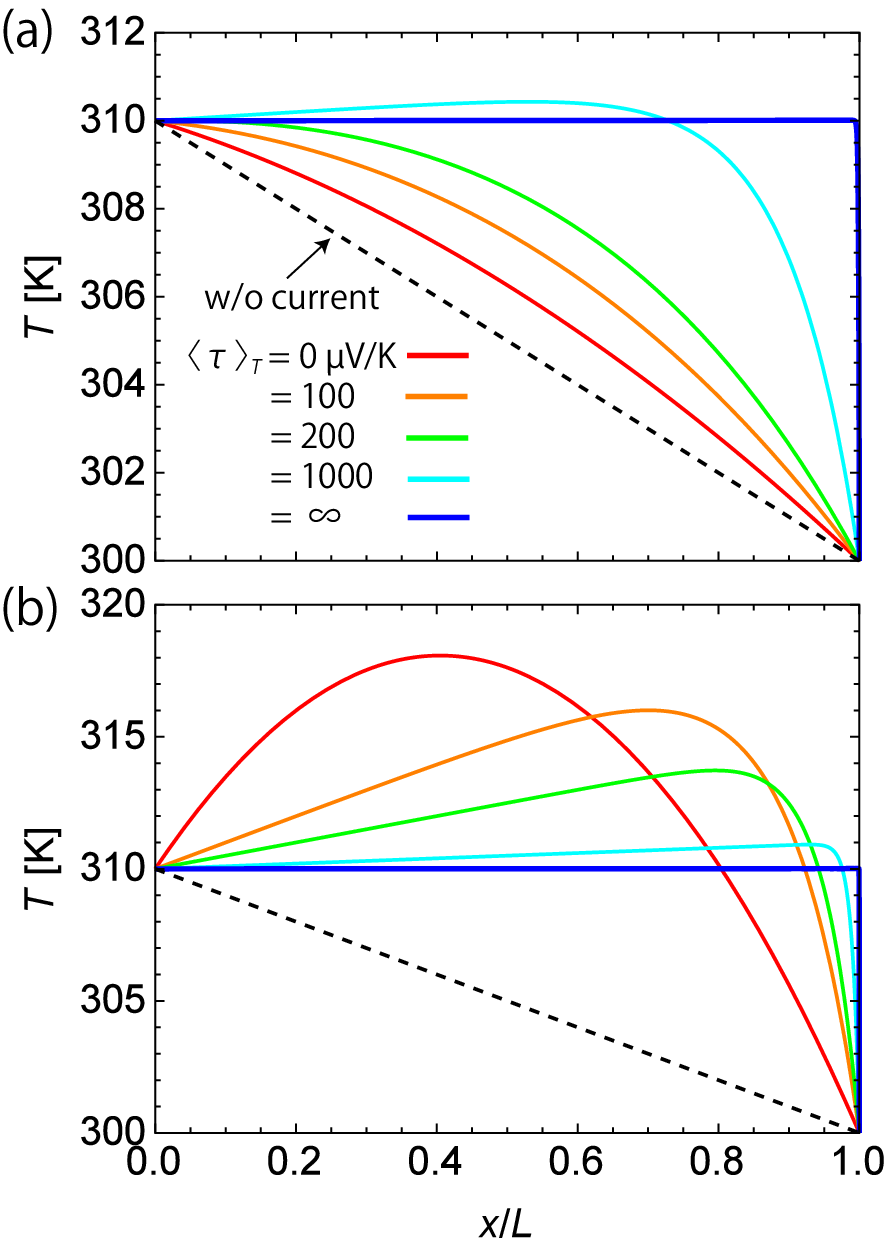} 
\par\end{centering}
\caption{Local temperature in the case of a heat release as a function of a normalized position $x/L$ for different values of $\langle \tau \rangle_T$~[$\umu$V/K] with (a) $K = 10$~$\umu$W/K and (b) $K = 1$~$\umu$W/K. The situation in the absence of current is described by dashed line. In these plots, $R = 0.01~\Omega$, $I = 0.1$~A, $T_{\rm c} = 300$~K, and $\Delta T = 10$~K are used. 
 }
\label{fig:T(x)h}
\end{figure}

Figures~\ref{fig:T(x)c} and \ref{fig:T(x)h} show a calculated local temperature $T(x)$ for different values of $\langle \tau \rangle_T$ and $K$. In these plots, we assume $R = 0.01~\Omega$, which is a reasonable value for conductors as seen in the next section. Figure~\ref{fig:T(x)c} corresponds to the Thomson-effect-induced heat absorption whereas Fig.~\ref{fig:T(x)h} is in the case of a heat release. For the cooling case of the Thomson-effect-induced bulk temperature modulation, the simplest example can be realized when for $\langle \tau \rangle_T < 0$ a charge current flows in the parallel direction of a temperature gradient.

\begin{figure}[ptb]
\begin{centering}
\includegraphics[width=0.45\textwidth,angle=0]{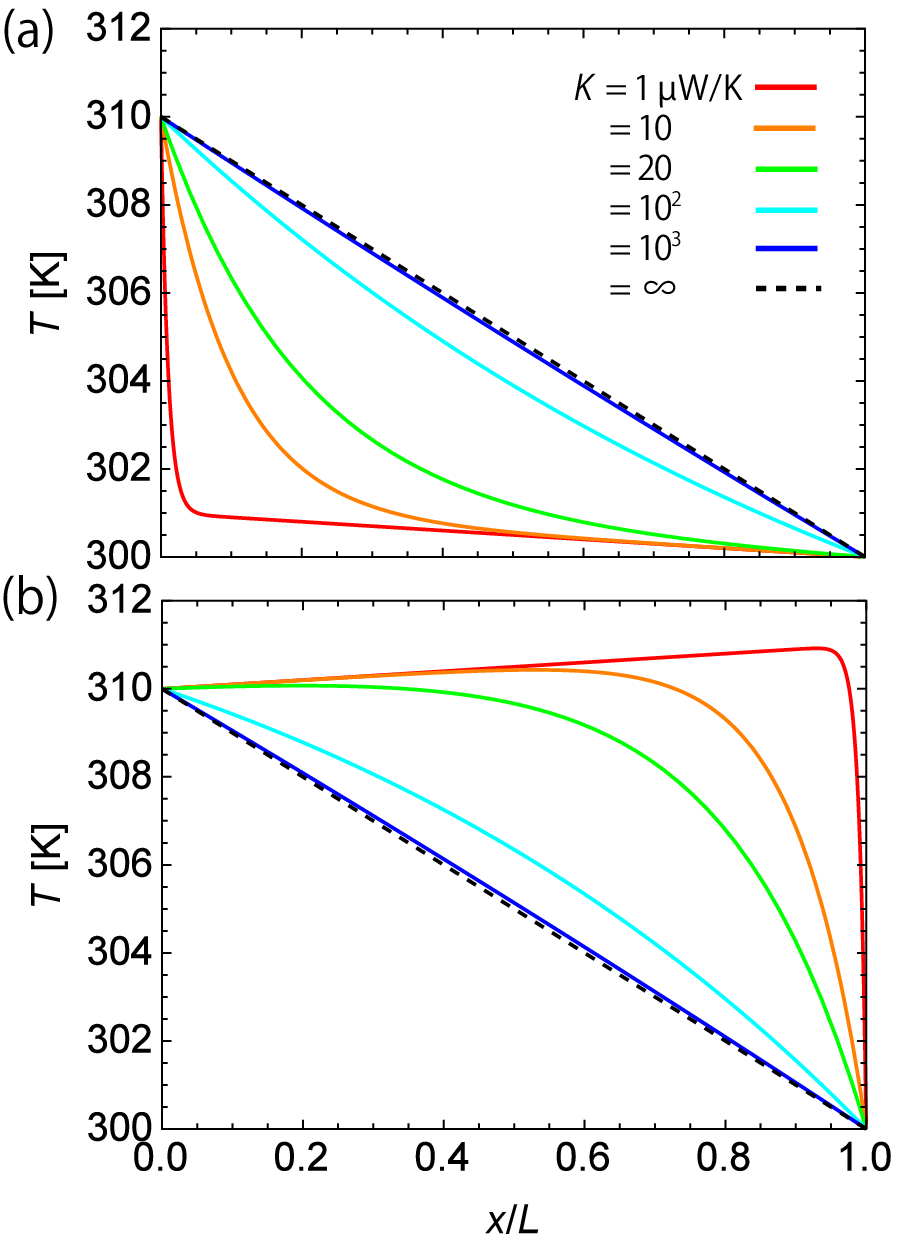} 
\par\end{centering}
\caption{Local temperature as a function of a normalized position $x/L$ for different values of $K$~[$\umu$W/K] in the case of (a) a heat absorption ($\langle \tau \rangle_T < 0$) and (b) a heat release ($\langle \tau \rangle_T > 0$). In these plots, $R = 0.01~\Omega$, $|\langle \tau \rangle_T| = 1000$~$\umu$V/K, $I = 0.1$~A, $T_{\rm c} = 300$~K, and $\Delta T = 10$~K are used. 
 }
\label{fig:T(x)K}
\end{figure}

First, we focus on the case with a small $|\langle \tau \rangle_T| \leq 100$~$\umu$V/K. Then, Eq.~(\ref{T(x)}) can be approximated by
\begin{align}
T(x) 
= T_h - \frac{RI^2}{2K}\left(  \frac{x}{L}\right)^2 
- \left(  \Delta T - \frac{RI^2}{2K}\right)\frac{x}{L}
,\label{T(x)2}
\end{align}
which is a well-known convex parabola distribution representing the Joule heating~(see both Figs.~\ref{fig:T(x)c} and \ref{fig:T(x)h}~(b)).
When the Joule heating is much smaller than the Fourier heat, Eq.~(\ref{T(x)2}) is further approximated by 
\begin{align}
T(x)
= T_{\rm h} - \Delta T\frac{x}{L}
,\label{Tgradient}
\end{align}
which describes an uniform temperature profile along the $x$-axis.
It is notable that regardless of the magnitude of $K$, Eq.~(\ref{Tgradient}) is always exact for $RI/\left(  \langle \tau \rangle_T\Delta T\right) = -1$~(corresponding to a case with $\langle \tau \rangle_T = -100$~$\umu$V/K in Fig.~\ref{fig:T(x)c}), which means that the Joule heating is cancelled out by the Thomson heat. In order to capture the influence of the thermal conductance on the temperature profile, we show the temperature profile for variation of $K$ in Fig.~\ref{fig:T(x)K}, in which one can see crossover from Eq.~(\ref{Ttau}) to Eq.~(\ref{Tgradient}) with increasing $K$.

For a large $|\langle \tau \rangle_T| \gg 100$~$\umu$V/K, the Thomson cooling overcomes the Joule heating (see Fig.~\ref{fig:T(x)c}) so that Eq.~(\ref{T(x)}) can be approximated by
\begin{align}
T(x)
= T_{\rm h} - \Delta T\frac{ \exp{\left(  \frac{\langle \tau \rangle_TI}{K}\frac{x}{L}\right)}- 1}{\exp{\left(  \frac{\langle \tau \rangle_TI}{K}\right)} - 1}
.\label{Ttau}
\end{align}
Note that $T(x) \geq T_{\rm c}$ for $\langle \tau \rangle_T \to -\infty$, which indicates that the minimum temperature in the system is determined by the cold-side heat bath due to a feedback of the Thomson effect. Even in the heating case as shown in Fig.~\ref{fig:T(x)h}~(b), the Thomson effect partially suppresses the Joule heating. With increasing $\langle \tau \rangle_T$, the maximum temperature in the system gradually coincides with that of the hot-side heat bath due to the feedback effect, i.e., $T(x) \to T_{\rm h}$ within $0 \leq x < L$ for $\langle \tau \rangle_T \to \infty$. These results reflect the fact that the Thomson-effect-induced heat production is dependent on the position and proportional to the local temperature gradient, which is self-consistently determined by the superposition of the applied temperature difference, Joule heating, and Thomson effect. For application of thermal management, the feedback effect might give a functionality that stabilizes the system temperature to one-side of the heat bath.

\subsection{System temperature difference $\Delta T_{\rm system}$}

\begin{figure}[ptb]
\begin{centering}
\includegraphics[width=0.45\textwidth,angle=0]{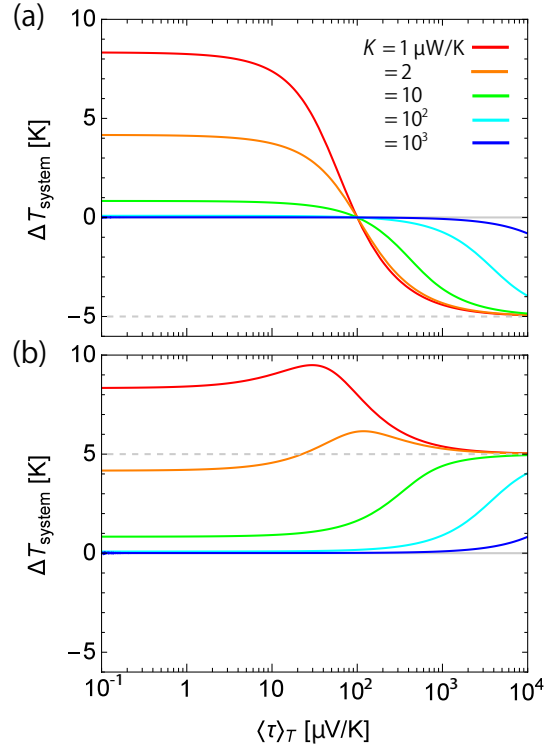} 
\par\end{centering}
\caption{System temperature difference as a function of $\langle \tau \rangle_T$~[$\umu$V/K] for various values of $K$~[$\umu$W/K]. (a) $I < 0$ for the Thomson-effect-induced heat absorption and (b) $I > 0$ for the case of a heat release. In this plot, $R = 0.01~\Omega$, $|I| = 0.1$~A, $T_{\rm c} = 300$~K, and $\Delta T = 10$~K are used. 
 }
\label{fig:DTsystem}
\end{figure}

To characterize the Thomson-effect-induced bulk temperature modulation, we introduce a system temperature difference $\Delta T_{\rm system}$ measured from $T_{\rm m} = \left(  T_{\rm h} + T_{\rm c}\right)/2$, which is defined by
\begin{align}
\Delta T_{\rm system} 
= \langle T \rangle_x - T_{\rm m}
,\label{DTsystem}
\end{align}
where 
\begin{align}
\langle T \rangle_x
= \frac{1}{L}\int_0^LT(x)dx
\label{Tbar}
\end{align}
is an averaged temperature across the conductor and $T_{\rm m}$ corresponds to $\langle T \rangle_x$ for the uniform temperature gradient (see Eq.~(\ref{Tgradient})). Hence, $\Delta T_{\rm system} < 0$ means that the system is cooled down from the initial state without the charge current by the Thomson effect. 

Calculated $\Delta T_{\rm system}$ in the case of heat absorption and release due to the Thomson effect is shown in Fig.~\ref{fig:DTsystem}. According to Eq.~(\ref{Thomson q}), we can control the heat release or absorption depending on a sign of a charge current (i.e. current reversal) for a constant $\langle \tau \rangle_T$. When we assume a conductor with $\langle \tau \rangle_T > 0$, $I < 0$ for the Thomson-effect-induced heat absorption and $I > 0$ for the case of a heat release. In the case with a small thermal conductance ($K < 10$~$\umu$W/K), the current reversal shows an asymmetry in the Thomson-effect-induced temperature modulation, which indicates that the Thomson effect partially suppresses the Joule heating even in the heat release.
As seen in Fig.~\ref{fig:DTsystem}, larger $\langle \tau \rangle_T$ and smaller $K$ are desirable for cooling the system temperature. Hence, for an efficient temperature modulation utilizing the Thomson effect, one will seek a material having larger $|\langle \tau \rangle_T|$ and smaller $K$. Taking into account this point, we perform further investigations in the next section by using realistic thermoelectric coefficients for various materials.

\subsection{Material requirements}

Here, we estimate the local temperature in the junctionless single conductor by using realistic thermoelectric coefficients for Bi$_{2}$Te$_{3}$ and FeRh-based alloy \cite{Modak21}. We also investigate an influence on the local temperature from the magneto-Thomson effect in BiSb alloy \cite{Uchida20}.

\subsubsection{Bi$_{2}$Te$_{3}$}

Bismuth telluride, Bi$_{2}$Te$_{3}$, is a commercial material typically used for the Peltier cooler and Seebeck power generator because Bi$_{2}$Te$_{3}$ has the top-level value of $ZT$ around room temperature \cite{Witting19}. Furthermore, a relatively large Thomson coefficient $\sim 150$~${\rm \umu V/K}$ at 300~K is predicted in a p-type Bi$_{2}$Te$_{3}$ by experiments \cite{Witting19} and first-principles calculation \cite{Hinsche16}. This Thomson coefficient could be even more enhanced for wide-gap semiconductors with very low carrier concentrations \cite{Seifert13,Hinsche16}. However, as seen in Table~\ref{tab.parameterBiTe}, the low carrier concentrations lead to increasing of electrical resistivity, which generates a huge Joule heating. Consequently, even with a relatively large Thomson coefficient, the heat absorption by the Thomson effect becomes negligible and Bi$_{2}$Te$_{3}$ is ineligible for cooling applications based on the Thomson effect. 

\begin{table}[ptb]
\begin{center}\caption{\label{tab.parameterBiTe}Parameters for $p$-type Bi$_2$Te$_3$ and Ni-doped Fe$_{49}$Rh$_{51}$.}
\begin{tabular}{lllll}
\hline
\multicolumn{1}{c}{} & \multicolumn{1}{c}{Symbol} & \multicolumn{1}{c}{\footnotemark[1]$^{-}$\footnotemark[3]$p$-type Bi$_2$Te$_3$} & \multicolumn{1}{c}{\footnotemark[4]Ni-doped Fe$_{49}$Rh$_{51}$} & \multicolumn{1}{c}{Unit} \\ \hline
Electrical resistivity & $\langle \rho \rangle_T$ & $7.5\times10^{-6}$ & $1.1\times10^{-6}$ & $\Omega$m \\
Seebeck coefficient & $\langle S \rangle_T$ & 170 & $-11$ & $\umu$VK$^{-1}$ \\
Thomson coefficient & $\langle \tau \rangle_T$ & 150 & $-906$ & $\umu$VK$^{-1}$ \\
Thermal conductivity & $\langle \kappa \rangle_T$ & 1.9 & 15 & ${\rm Wm^{-1}K^{-1}}$ \\ \hline
Electrical resistance & $R$ & 0.14 & 0.019 & $\Omega$ \\
Thermal conductance & $K$ & 1.0$\times10^2$ & 8.2$\times10^2$ & ${\rm \umu WK^{-1}}$ \\  
System temperature difference & $\Delta T_{\rm system}$ & 10 & $-0.10$ & K \\ \hline
Joule heat & $RI^2$ & 13 & $1.7$ & mW \\
Thomson heat & $\langle \tau \rangle_T\Delta TI$ & $-4.5$ & $-2.7$ & mW \\  
Fourier heat & $K\Delta T$ & $1.0$ & $8.2$ & mW \\ \hline
\end{tabular}
\end{center}
\footnotesize{~~~\footnotemark[1]{Reference~\citen{Witting19},}
\footnotemark[2]{Reference~\citen{Hinsche16},}
\footnotemark[3]{Reference~\citen{Bohra18},}
\footnotemark[4]{Reference~\citen{Modak21},}
\footnotemark[5]{Other parameters: $\Delta T = 10$~K, $A \approx 0.77$~mm$^2$, $L = 14$~mm, $|I| \approx 0.3$~A.}}
\end{table}

\begin{figure}[ptb]
\begin{centering}
\includegraphics[width=0.45\textwidth,angle=0]{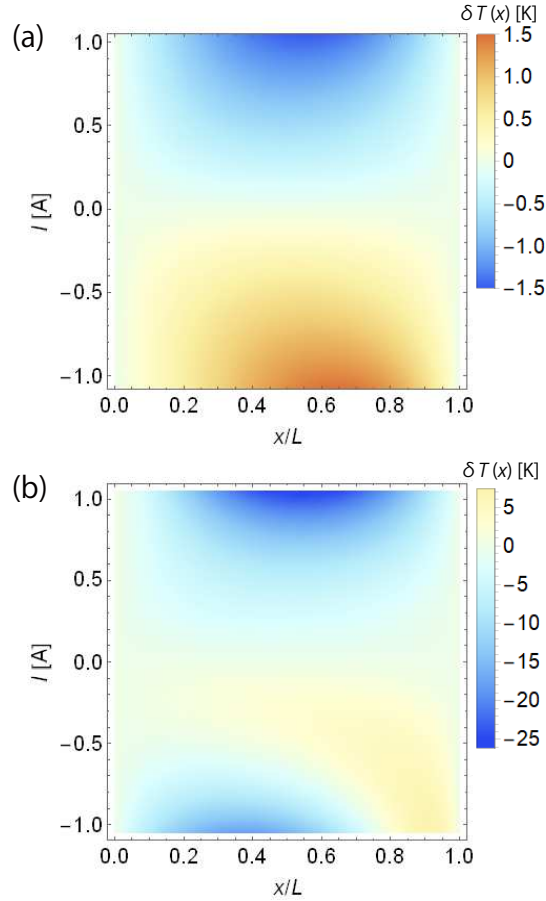} 
\par\end{centering}
\caption{Local temperature difference $\delta T(x)$ in a FeRh-based alloy as a function of a normalized position $x/L$ and a charge current $I$ for (a) $K = 8.2\times10^2$~${\rm \umu W/K}$ and (b) $\langle \kappa \rangle_T = 82$~${\rm \umu W/K}$~(assumption). In these plots, $T_{\rm c} = 312$~K and $\Delta T = 10$~K are used. The other parameters are listed in Tables~\ref{tab.parameterFeRh}. The definition of $\delta T(x)$~(see Eq.~(\ref{dT})) is explained in the text.
 }
\label{fig:T(x)MR}
\end{figure}

\subsubsection{FeRh-based alloy}

FeRh-based alloys are well-studied magnetic materials that exhibit a first-order antiferromagnetic-ferromagnetic phase transition around room temperature \cite{Kudrnovsky15}. This phase transition accompanies a steep $T$ dependence of the transport properties \cite{Algarabel95,Mankovsky17}. Recently, due to the steep $T$ dependence of the Seebeck coefficient, it is found that Ni-doped Fe$_{49}$Rh$_{51}$ with a magnetization $\umu_0M \approx 1.3$~T show huge values of the Thomson coefficient approaching $\sim 1000$~$\umu$V/K around room temperature \cite{Modak21}.
To elucidate how the Thomson effect modifies the local temperature, we define a local temperature difference in the presence/absence of the Thomson effect
\begin{align}
\delta T(x)
= T(x) - T(x)|_{\tau = 0}
,\label{dT}
\end{align}
where $T(x)|_{\tau = 0}$ denotes the local temperature in the absence of the Thomson effect.
Figure~\ref{fig:T(x)MR} shows calculated $\delta T(x)$ for a Ni-doped Fe$_{49}$Rh$_{51}$ with material parameters listed in Table~\ref{tab.parameterBiTe}. Note that in actual the phase-transition-induced giant Thomson coefficient in Table~\ref{tab.parameterFeRh} is obtained within a narrow temperature range (a several Kelvin) but for simplicity we assume keeping the highest value in our calculation. In addition, the FeRh-based alloy has a negative Thomson coefficient $\langle \tau \rangle_T < 0$. As seen in Fig.~\ref{fig:T(x)MR}~(a), the Thomson-effect-induced cooling (for $I > 0$) and heating (for $I < 0$) become maxim around the center ($x/L = 0.5$) of the device. For $I > 0$, although the FeRh-based alloy has a relatively high electrical resistivity compared with pure metals, the Joule heating is suppressed by the Thomson-effect-induced cooling. Consequently, the system with a current is cooled down and obtains $\Delta T_{\rm system} = -0.10$~K. As shown in this example, the Thomson-effect-induced cooling in the bulk region is possible even when the Seebeck coefficient are small, which is also confirmed by a recent experiment \cite{Modak21}. If we could reduce the thermal conductivity down to $1.5$~${\rm Wm^{-1}K^{-1}}$ without changing other parameters, the system is farther cooled down and obtains $\Delta T_{\rm system} = -0.83$~K. Corresponding temperature modulation by the Thomson effect is shown in Fig.~\ref{fig:T(x)MR}~(b). As seen, current reversal shows an asymmetry in the Thomson-effect-induced temperature modulation and the cooling case is strongly enhanced. Even in the heating case, the Thomson effect partially suppresses the Joule heating. 

\begin{figure}[ptb]
\begin{centering}
\includegraphics[width=0.45\textwidth,angle=0]{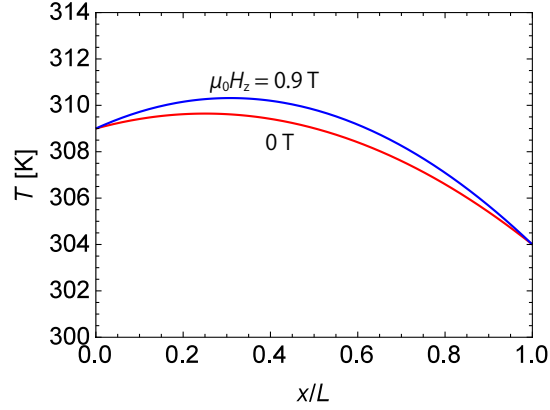} 
\par\end{centering}
\caption{Local temperature as a function of a normalized position $x/L$ in a BiSb alloy in the presence/absence of a magnetic field $\umu_0 H_z$. In this plot, $T_{\rm c} = 304$~K and $\Delta T = 5$~K are used. The other parameters are listed in \ref{tab.parameterBiSb}.
 }
\label{fig:T(x)MR1}
\end{figure}

\begin{table}[ptb]
\begin{center}\caption{\label{tab.parameterBiSb}Parameters for Bi$_{88}$Sb$_{12}$ in a magnetic field $\umu_0H_z$~[T].}
\begin{tabular}{lllll}
\hline
\multicolumn{1}{c}{} & \multicolumn{1}{c}{Symbol} & \multicolumn{1}{c}{\footnotemark[1]$\umu_0H_z = 0$~T} & \multicolumn{1}{c}{\footnotemark[1]$\umu_0H_z = 0.9$~T} & \multicolumn{1}{c}{Unit} \\ \hline
Electrical resistivity & $\langle \rho \rangle_T$ & $3.1\times10^{-6}$ & $3.8\times10^{-6}$ & $\Omega$m \\
Seebeck coefficient & $\langle S \rangle_T$ & $-91$ & $-110$ & $\umu$VK$^{-1}$ \\
Thomson coefficient & $\langle \tau \rangle_T$ & 45 & 98 & $\umu$VK$^{-1}$ \\Thermal conductivity & $\langle \kappa \rangle_T$ & 3.5 & 3.2 & ${\rm Wm^{-1}K^{-1}}$ \\ \hline 
Electrical resistance & $R$ & 0.15 & 0.18 & $\Omega$ \\
Thermal conductance & $K$ & 72 & 67 & ${\rm \umu WK^{-1}}$ \\  
System temperature difference & $\Delta T_{\rm system}$ & 1.7 & 2.3 & K \\ \hline
Joule heat & $RI^2$ & $1.5$ & $1.8$ & mW \\
Thomson heat & $\langle \tau \rangle_T\Delta TI$ & $-0.045$ & $-0.098$ & mW \\  
Fourier heat & $K\Delta T$ & $0.72$ & $0.67$ & mW \\ \hline
\end{tabular}
\end{center}
\footnotesize{~~~~~~~~~~~~~~~~~~\footnotemark[1]{Reference~\citen{Uchida20},}
\footnotemark[2]{Other parameters: $\Delta T = 10$~K, $A = 0.25$~mm$^2$, $L = 12$~mm, $I \approx -0.1$~A.}}
\end{table}

\subsubsection{BiSb alloy}%

BiSb alloys (Bi$_{1-x}$Sb$_{x}$) show various solid-state phases such as semimetal, narrow-gap semiconductor, and three-dimensional topological insulator by tunning the composition $x$ \cite{Hsieh08}. The thermomagnetic properties of BiSb alloys have also been studied intensively so far \cite{Yim97,Murata16}. In 2020, Uchida \textit{et.~al.} 
reported that the Thomson coefficient of Bi$_{88}$Sb$_{12}$ is strongly enhanced by applying a magnetic field, as shown in Table.~\ref{tab.parameterBiSb}. 
Here, we investigate an influence on the local temperature from the magneto-Thomson effect in a BiSb alloy. 
Figure~\ref{fig:T(x)MR1} shows calculated local temperature for Bi$_{88}$Sb$_{12}$ with material parameters listed in Table~\ref{tab.parameterBiSb}. In the presence of a magnetic filed, the electrical resistivity increases due to the magnetoresistance effect, leading to an enhancement of Joule heating. Note that the Thomson-effect-induced cooling is also enhanced by the magneto-Thomson effect but is less than that of the Joule heating. Consequently, the system temperature becomes higher than that of zero magnetic field case and obtains $\Delta T_{\rm system} = 2.3$~K. Therefore, for an efficient cooling owing to the magneto-Thomson effect, materials with a larger magnetic field dependence of the Thomson coefficient will be required. In this sense, three-dimensional topological insulator phase of BiSb alloys \cite{Hsieh08} and Bi$_{2}$Te$_{3}$ tetradymite family \cite{Witting19} might be potential candidates for the magneto-Thomson effect because of a drastic transport property change on a topological phase transition induced by a magnetic field \cite{Satake20,Chiba19}. 

\section{Summary}\label{Summary}

In summary, we have theoretically investigated the temperature profile due to the Thomson-effect-induced heat release/absorption in junctionless single conductors. Under the simultaneous application of a charge current and a temperature gradient to the conductor, the system can be a simple temperature modulator without constructing heterojunction structure. For a conductor having a large Thomson coefficient, we showed that the Thomson-effect-induced heat absorption overcomes the Joule heating, resulting in current-induced cooling in the bulk region from the initial state without the charge current. Even in the heating case with a small thermal conductance, the Thomson effect partially suppresses the Joule heating, i.e., current reversal shows an asymmetry in the Thomson-effect-induced temperature modulation. We also revealed that the feedback of the Thomson effect stabilizes the system temperature to one-side of the heat bath, which reflects the fact that the Thomson-effect-induced heat production is directly proportional to the local temperature gradient. The feedback effect might give a new functionality for thermal management technology if materials with far greater Thomson coefficients than currently exist can be found.
Moreover, we estimate the local temperature of the junctionless single conductor with realistic thermoelectric coefficients. It is demonstrated that the system with a current can be cooled down by using parameters of a FeRh-based alloy. We have also investigated an influence on the local temperature from the magneto-Thomson effect in a BiSb alloy. This work will be the basis for designing the Thomson-effect-based thermoelectric devices and for finding materials for nonlinear thermoelectrics and spin caloritronics.

\begin{acknowledgment}


The authors thank Rajkumar Modak, Jun'ichi Ieda, Yasufumi Araki, Kei Yamamoto, and Gerrit. E. W. Bauer for valuable discussions.
This work was supported by Grants-in-Aid for Scientific Research (Grant Nos.~ 20K15163, 20H02196, 21K18590, and 19H02585) from JSPS KAKENHI, Japan, and CREST "Creation of Innovative Core Technologies for Nano-enabled Thermal Management" (Grant No.~JPMJCR17I1) from JST, Japan. 

\end{acknowledgment}


\end{document}